\documentclass[11pt]{article}
\usepackage{epsfig}
\usepackage{amsfonts}
\usepackage{amsmath}
\usepackage{amssymb}
\usepackage{color}

\newcommand{\w}{\omega}

\newcommand{\al}{\alpha}
\newcommand{\bt}{\beta}

\newcommand{\slashP}{\slash\hspace{-0.6em}P}
\newcommand{\slashPsub}{\slash\hspace{-0.5em}P}

\newcommand{\Nb}{\bar N}
\newcommand{\Fp}{F_\pi}

\newcommand{\mpi}{m_{\pi}}

\newcommand{\Or}{\mathcal O}

\newcommand{\dslash}[1]{#1 \llap{/\kern-0.5pt}}
\newcommand{\Dslash}[1]{#1 \llap{/\kern+1.2pt}}
\newcommand{\DDslash}[1]{#1 \llap{/\kern+2.3pt}}
\newcommand{\dslashh}[1]{#1 \llap{/\kern+1pt}}

\newcommand{\bea}{\begin{eqnarray}}
\newcommand{\eea}{\end{eqnarray}}
\newcommand{\bma}{\begin{pmatrix}}
\newcommand{\ema}{\end{pmatrix}}

\begin{document}
\begin{titlepage}

\vspace{2.0cm}

\begin{center}
{\Large\bf
A study of the parity-odd nucleon-nucleon potential}
\vspace{1.7cm}

{\large \bf   J. de Vries$^{1}$, N. Li$^{1}$, Ulf-G. Mei{\ss}ner$^{1,2}$,
  N. Kaiser$^{3}$, X.-H. Liu$^{4,5}$, S.-L. Zhu$^{4,5}$}

\vspace{0.5cm}

{\large
$^1$
{\it Institute for Advanced Simulation, Institut f\"ur Kernphysik,
and J\"ulich Center for Hadron Physics, Forschungszentrum J\"ulich,
D-52425 J\"ulich, Germany}}

\vspace{0.25cm}
{\large
$^2$
{\it Helmholtz-Institut f\"ur Strahlen- und Kernphysik and Bethe Center for
Theoretical Physics, Universit\"at Bonn, D-53115 Bonn, Germany}}

\vspace{0.25cm}
{\large
$^3$
{\it Physik Department T39, Technische Universit\"at M\"unchen, D-85747 Garching, Germany}}

\vspace{0.25cm}
{\large
$^4$
{\it
Collaborative Innovation Center of Quantum Matter, Beijing 100871,
China} }

\vspace{0.25cm}
{\large
$^5$
{\it
Department of Physics and State Key Laboratory of Nuclear
Physics and Technology, Peking University,
Beijing 100871, China
} }

\end{center}

\vspace{1.5cm}

\begin{abstract}
We investigate the parity-violating nucleon-nucleon potential as obtained in chiral effective field theory. By using resonance saturation we compare the chiral potential to the more traditional one-meson exchange potential. In particular, we show how parameters appearing in the different approaches can be compared with each other and demonstrate that analyses of parity violation in proton-proton scattering within the different approaches are in good agreement. In the second part of this work, we extend the parity-violating potential to next-to-next-to-leading order. We show that generally it includes both one-pion- and two-pion-exchange corrections, but the former play no significant role. The two-pion-exchange corrections depend on five new low-energy constants which only become important if the leading-order weak pion-nucleon constant $h_\pi$ turns out to be very small.
\end{abstract}

\vfill
\end{titlepage}

\section{Introduction}
Despite decades of experimental and theoretical effort hadronic flavor-conserving parity ($P$) violation is still not well understood. At low energies, the Standard Model Lagrangian contains parity-violating ($\slashP$) four-fermion operators which arise when the heavy weak bosons decouple. The charged-current operators lead to beta-decay of leptons, neutrons, and nuclei, while the neutral-current operators give rise to, among other processes, neutrino-nucleus scattering and $\slashP$ hadronic interactions. A satisfying description of the latter is currently missing, a problem which is mainly caused by the breakdown of perturbation theory at low energies and the high experimental accuracy required to measure parity-violating observables in hadronic and nuclear systems.

Hadronic $P$ violation has so far only been measured in a handful of processes while upper bounds have been obtained in other experiments. Finite signals were obtained for several energies in proton-proton ($pp$)~\cite{Kistryn:1987tq,Eversheim:1991tg,Berdoz:2001nu} and $p\alpha$ scattering~\cite{Henneck:1982cc, Lang:1985jv}, radiative decays of ${}^{19}$F~\cite{Elsener:1984vp,Elsener:1987sx}, and the anapole moment of the Cesium atom~\cite{Wood:1997zq}. Strong upper bounds are found for $np$ capture~\cite{Cavaignac:1977uk, Gericke:2011zz}, radiative decays of ${}^{18}$F~\cite{Adelberger:1983zz,Adelberger:1983zz2}, and several other observables. For recent reviews we refer to Refs.~\cite{Holsteinreview,Schindler:2013yua}.

In order to interpret the data, $P$ violation in hadronic systems has often been parametrized by a one-meson exchange model. In this model, usually called the DDH model after the authors of Ref.~\cite{Desplanques:1979hn}, $P$ violation is induced by the exchange of a single pion, $\rho$-, or $\omega$-meson between two nucleons. The resulting $\slashP$ nucleon-nucleon $(N\!N)$ potential depends on seven weak meson-nucleon coupling constants and can be combined with phenomenological $P$-even $N\!N$ potentials to calculate $\slashP$ observables.

A more systematic description of hadronic $P$ violation can be obtained by
using chiral effective field theory ($\chi$EFT), the low-energy effective field
theory (EFT) of QCD. $\chi$EFT allows for the systematic construction of the
interactions among pions and nucleons and, in principle, heavier baryons. For
reviews on $\chi$EFT, see e.g. Refs.~\cite{Bernard:2006gx,  Epelbaum:2008ga, Machleidt:2011zz}. At leading order (LO) the Lagrangian consists of a single $\slashP$ pion-nucleon interaction with coupling constant $h_\pi$ \cite{Kaplan:1992vj} which induces, in combination with the standard pseudovector $P$-even pion-nucleon interaction, the LO $\slashP$  $N\!N$ potential. At next-to-leading order (NLO), the Lagrangian consists of five additional $\slashP$  $N\!N$ contact interactions \cite{Savage:1998rx, Kaplan:1998xi, Savage:2000iv, Zhu, Girlanda:2008ts} which give rise to a short-range $\slashP$ potential. At the same order, two-pion-exchange (TPE) contributions to the potential appear which, just as the LO potential, depend solely on $h_\pi$ \cite{Zhu, KaiserPodd}.

The application of $\chi$EFT to describe hadronic $P$ violation has a few advantages over the more traditional DDH model. First of all, there is a clear link to the underlying theory, \textit{i.e.}, QCD supplemented with $\slashP$ four-quark operators. Second, the $\chi$EFT approach makes it possible to calculate the $P$-even and -odd $N\!N$ potentials within the same framework. The resulting potentials can then be treated on the same footing. This allows, for example, for the systematic variation of the cut-offs appearing in the solution of scattering equations which gives an estimate of the theoretical errors in the calculation \cite{deVries, Viviani}. Third, the chiral Lagrangian can be improved by going to higher orders in the expansion. In fact, in this paper we study next-to-next-to-leading-order (N$^2$LO) corrections to the $\slashP$ potential. Fourth, the chiral approach can be extended to other systems, such as reactions with more than two nucleons \cite{Viviani, Griesshammer:2010nd, Griesshammer:2011md}, or processes involving photons \cite{Savage:1998rx, Kaplan:1998xi,Savage:2000iv}. The latter require the calculation of $\slashP$ currents. These currents can be evaluated within the same framework as the potential, something which is not possible in the DDH model where the currents need to be modeled separately.

The task is then to use the $\slashP$ potential to calculate $\slashP$ effects in processes such as $N\!N$ scattering, nuclear break-up or capture reactions, and nuclear anapole moments. By comparison with data, the $\slashP$ low-energy constants (LECs) can be extracted. The values of the LECs can then be used to predict other processes. In addition, the extracted values can be compared with model \cite{ Desplanques:1979hn, Dubovik:1986pj, Kaiser1, Feldman:1991tj, Meissner1} and lattice calculations \cite{hpilatt} of the LECs, which provides a handle to judge various calculational methods.

Most works have used the DDH potential instead of the chiral effective potential. It would therefore be useful to be able to compare results obtained using the two potentials. At very low energies all mesons can be integrated out and both approaches collapse to a potential consisting of \mbox{$\slashP$} contact interactions between nucleons. In this energy range, the two approaches become identical and the different LECs can be easily compared \cite{Phillips:2008hn, Holstein:2009zzb}. At higher energies, the heavier mesons can still be integrated out, but the pion becomes dynamical which makes the comparison more complicated. Both the chiral effective and the DDH potential contain a one-pion-exchange contribution proportional to $h_\pi$, but the chiral potential contains also TPE corrections which need to be taken into account. In this work we use resonance saturation techniques \cite{ Epelbaum:2001fm, Berengut:2013nh} to derive a dictionary between the two approaches.

The experimental upper bounds on $P$ violation in $np$ capture \cite{Cavaignac:1977uk, Gericke:2011zz} and $\gamma$-ray emission of ${}^{18}$F \cite{Adelberger:1983zz, Adelberger:1983zz2}, set a rather strong bound on the size of $h_\pi$. Such a small value was also found in several model estimates \cite{Kaiser1, Meissner1} and recently by the first lattice-QCD calculation~\cite{hpilatt}. In recent work \cite{deVries}, we have investigated whether this smallness of $h_\pi$ is consistent with data on the longitudinal analyzing power (LAP) in $pp$ scattering. The LO $\slashP$ potential does not contribute to $pp$ scattering, however, via the NLO TPE contributions, there is still a dependence on $h_\pi$. Unfortunately, the lack of data forced us to adopt a rather large allowed range of $h_\pi$ and we could neither confirm nor rule out the small values of $h_\pi$. This conclusion was confirmed in Ref.~\cite{Viviani}. Additional data is needed to make statements about the size of $h_\pi$.

The suggested smallness of $h_\pi$ implies that higher-order corrections to the $\slashP$ $N\!N$ potential might be relevant.
In this work, we investigate these corrections by calculating, for the first time, the $\slashP$ potential up to next-to-next-to-leading order (N${}^2$LO). We study the size of these corrections by calculating their contributions to the $pp$ longitudinal analyzing power. If future data cannot be fitted by the six LECs appearing in the NLO potential, it might be that the N${}^2$LO corrections need to be taken into account.

This paper is divided into two main parts. In Sec.~\ref{resonance} we review the NLO $\slashP$ chiral potential and the DDH potential. We use resonance saturation techniques to express the LECs appearing in the chiral potential in terms of the DDH parameters. Some details are given in Appendix~\ref{AppA}. We then compare the two different frameworks using data on $pp$ LAP. The second part of the paper starts in Sec.~\ref{N2LOpot} where we, motivated by the possible smallness of $h_\pi$, calculate the N${}^2$LO corrections to the $\slashP$ potential. We study the size of these corrections by calculating their contributions to the $pp$ LAP. We conclude in Sect.~\ref{conclusion}.

\section{The parity-odd nucleon-nucleon potential in different frameworks}\label{resonance}
\subsection{The chiral parity-odd potential up to next-to-leading order}
At leading order (LO) in the power counting, the only term appearing in the chiral Lagrangian is the weak pion-nucleon vertex
\begin{equation}\label{PoddLO}
\mathcal L_{\slashPsub } = \frac{h_\pi}{\sqrt{2}} \Nb (\vec \pi\times \vec \tau)^3 N\,\,\,,
\end{equation}
proportional to the LEC  $h_\pi$ \cite{Kaplan:1992vj}. In this expression, $N= (p, n)^t$ denotes the nucleon isospin-doublet, $\vec \pi$ the pion isospin-triplet, and $\vec \tau$ the isospin Pauli
matrices. Together with the usual pseudovector $P$-conserving
pion-nucleon interaction, the LO $\slashP$ OPE potential follows as
\begin{equation}
V_{\text{OPE}}
= - \frac{g_{A}h_\pi}{ 2\sqrt{2} F_\pi} i(\vec \tau_1\times \vec \tau_2)^3 \frac{(\vec \sigma_1+\vec \sigma_2)\cdot \vec q }{\mpi^2+q^2}\,\,\,,
\label{onepion}
\end{equation}
in terms of the nucleon spin $\vec \sigma_{1,2}$ and the momentum transfer flowing from nucleon $1$ to nucleon $2$: $\vec q = \vec p - \vec p^{\,\prime}$ ($q = |\vec q\,|$), where $\pm \vec p$ and $\pm \vec
p^{\,\prime}$ are the momenta of the incoming and
outgoing nucleons in the center-of-mass frame. Other parameters appearing in Eq.~\eqref{onepion} are the pion decay constant $\Fp = 92.4$\,MeV, the charged pion-mass $\mpi = 139.57$\,MeV, and the nucleon axial-vector coupling constant $g_A=1.29$ (taking into account the Goldberger-Treiman discrepancy).
The LO OPE potential changes the total isospin of the interacting nucleon pair and, at low energies, dominantly contributes to the ${}^3 S_1\leftrightarrow {}^3 P_1$ transition. The isospin change ensures that the LO potential vanishes for $pp$ and $nn$ scattering.

At NLO the number of LECs proliferates. First of all, $N\!N$ short-range contact interactions appear. These can be parametrized in many ways \cite{Girlanda:2008ts} and we choose the following form for the associated potential
\begin{eqnarray}\label{CT}
V_{\text{CT}}
&=& \frac{C_0}{ \Fp \Lambda_\chi^2}  (\vec \sigma_1 - \vec \sigma_2)\cdot (\vec p + \vec p^{\,\prime} )\nonumber\\
&& + \frac{1}{ \Fp \Lambda_\chi^2} \left(C_1 + C_2\frac{(\vec\tau_1+\vec\tau_2)^3}{2} + C_3\frac{\vec \tau_1\cdot \vec \tau_2-3 \tau_1^3 \tau_2^3 }{2} \right) i(\vec \sigma_1 \times \vec \sigma_2) \cdot \vec q\nonumber\\
&& + \frac{C_4}{ \Fp \Lambda_\chi^2} i(\vec \tau_1\times \vec \tau_2)^3(\vec \sigma_1 + \vec \sigma_2)\cdot \vec q\,\,\,,
\end{eqnarray}
where the factor $( \Fp \Lambda_\chi^2)^{-1}$ (with $\Lambda_\chi = 1$ GeV) is inserted to make the couplings $C_i$ dimensionless. In addition it ensures that the naive dimensional estimates\footnote{More detailed estimates which take into account factors of the Weinberg angle and strangeness effects can be found in Ref.~\cite{Kaplan:1992vj}.} of $h_\pi$ and $C_i$ are of the same size
\begin{equation}
h_\pi \sim C_i \sim \mathcal O(G_F \Fp \Lambda_\chi) \sim 10^{-6}\,\,\,.
\end{equation}
The contact part of the potential contains five independent LECs which can be understood from the fact that five different $S\leftrightarrow P$ transitions are possible.

Two-pion-exchange (TPE) diagrams appear at this order as well. These diagrams
involve LO vertices only, that is $h_\pi$ for the $\slashP$ vertex and $g_A$
or the Weinberg-Tomazawa interaction for the $P$-conserving vertices. The
relevant diagrams have been calculated several times \cite{Zhu, KaiserPodd}
and here we give the results using spectral function regularization \cite{Epelbaum:2003gr, deVries}
\begin{eqnarray}\label{NLOTPE}
V_{\text{TPE}}(q,\Lambda_S)
&=&   -\frac{ g_A h_\pi}{2 \sqrt{2}\Fp} \frac{1}{(4\pi \Fp)^2} i (\vec \tau_1\times \vec \tau_2)^3 (\vec \sigma_1+\vec \sigma_2)\cdot \vec q \left(g_A^2 \frac{8\mpi^2+3q^2}{\omega^2} - 1 \right) L(q,\Lambda_S)\nonumber\\
&& +  \frac{ g_A^3 h_\pi}{2 \sqrt{2}\Fp}  \frac{4}{(4\pi \Fp)^2} i (\vec \tau_1+ \vec \tau_2)^3 (\vec \sigma_1\times \vec \sigma_2)\cdot \vec q \,\, L(q, \Lambda_S)\,\,\,,
\end{eqnarray}
in terms of the loop functions
\begin{equation}
L(q,\Lambda_S)= \frac{\w}{2q} \log\frac{\Lambda_S^2 \w^2 +q^2 s^2 + 2 \Lambda_S s \w q}{4\mpi^2(\Lambda_S^2+q^2)}\,\,,\hspace{3mm}\w = \sqrt{q^2+4\mpi^2}\,\,, \hspace{3mm} s = \sqrt{\Lambda_S^2-4\mpi^2}\,\,\,.
\end{equation}

The contact interactions and TPE diagrams complete the potential up to NLO. In principle, one might expect additional contributions via corrections to the OPE potential. However, such corrections can either be absorbed into the LO LECs, vanish, or appear at higher order.

\subsection{The DDH potential}
Historically, the most frequently applied approach to hadronic $P$ violation is the one-meson exchange model (often called the DDH model, after the authors of Ref.~\cite{Desplanques:1979hn}). In this model, $P$ violation is induced due to the exchange of a single pion, $\rho$-, or $\omega$-meson. The exchange of a pion in the DDH model gives rise to the same potential\footnote{Note that sometimes a regulator function (similar to those in Eq.~\eqref{DDHcutoff}) is applied to the OPE potential in the DDH framework. } as Eq.~\eqref{onepion}. The  $\slashP$ potential in momentum space due to the exchange of the heavier mesons is given by
\begin{eqnarray}\label{DDHpot}
V_{\mathrm{DDH}}&=& \bigg\{ -\frac{ g_\rho}{m_N}\left[  \vec \tau_1\cdot \vec\tau_2  \,h_\rho^0
 + \frac{(\vec \tau_1+\vec \tau_2)^3}{2}  h_\rho^1 +\frac{3 \tau_1^3 \tau_2^3-\vec \tau_1\cdot \vec \tau_2}{2\sqrt 6} h_\rho^2\right] f_\rho(q^2)\nonumber\\
 &&-\frac{ g_\omega}{m_N}\left[h_\omega^0+ \frac{(\vec \tau_1+\vec \tau_2)^3}{2}  h_\omega^1 \right] f_\omega(q^2) \bigg\} (\vec \sigma_1 - \vec \sigma_2)\cdot (\vec p + \vec p^{\,\prime} )\nonumber\\
 && + \bigg\{ \frac{ g_\rho(1+\chi_V)}{m_N}\left[  \vec \tau_1\cdot \vec\tau_2 \, h_\rho^0
 + \frac{(\vec \tau_1+\vec \tau_2)^3}{2}  h_\rho^1 +\frac{3 \tau_1^3 \tau_2^3 -\vec \tau_1\cdot \vec \tau_2 }{2\sqrt 6} h_\rho^2\right] f_\rho(q^2)\nonumber\\
 &&+\frac{ g_\omega(1+\chi_S)}{m_N}\left[h_\omega^0+ \frac{(\vec \tau_1+\vec \tau_2)^3}{2}  h_\omega^1 \right] f_\omega(q^2) \bigg\}i (\vec \sigma_1 \times \vec \sigma_2)\cdot \vec q\nonumber\\
 && +\left[\frac{g_\rho h_\rho^1}{m_N}f_\rho(q^2)-\frac{g_\omega h_\omega^1}{m_N}f_\omega(q^2) \right]\frac{(\vec \tau_1-\vec \tau_2)^3}{2}(\vec \sigma_1 + \vec \sigma_2)\cdot (\vec p + \vec p^{\,\prime} )\nonumber\\
 && +\left[ \frac{g_\rho h_\rho^{1\,\prime}}{2 m_N} f_\rho(q^2)\right]i(\vec \tau_1 \times \vec \tau_2)^3(\vec \sigma_1 + \vec \sigma_2)\cdot \vec q\,\,\,,
\end{eqnarray}
in terms of the $P$-even vertices\footnote{Sometimes different values for the strong couplings are used which affects the extraction of the $\slashP$ parameters, see the discussion in Ref.~\cite{Holsteinreview}} $g_\omega =8.4 $, $g_\rho = 2.8$, $\chi_S=-0.12$, and $\chi_V = 3.70$, six $\slashP$ meson-nucleon couplings $h_\rho^{0,1,2}$, $h_\rho^{1\,\prime}$, and $h_\omega^{0,1}$, and the one-meson exchange functions
\begin{equation}
f_\rho(q^2) = \frac{1}{m_\rho^2 + q^2} \, c_\rho(q^2,\Lambda_\rho)\,\,\,,\qquad f_\omega(q^2) = \frac{1}{m_\omega^2 + q^2}\, c_\omega(q^2,\Lambda_\omega)\,\,\,,
\end{equation}
where $m_\rho \simeq m_\omega \simeq 780$ MeV are, respectively, the masses of the $\rho$- and $\omega$-meson. The functions $c_\rho$ and $c_\omega$ are cut-off functions which regulate the potentials. They are, however, not always applied and also their form can vary. We will use the following regulator functions which were used, for example, in Ref.~\cite{Carlson:2001ma}
\begin{equation}\label{DDHcutoff}
c_{\rho,\omega}(q^2,\Lambda_{\rho,\omega}) = \left(\frac{\Lambda_{\rho,\omega}^2 - m_{
\rho,\omega}^2}{\Lambda^2_{\rho,\omega}+q^2} \right)^2\,\,\,,
\end{equation}
in terms of the cut-off masses $\Lambda_\rho$ and $\Lambda_\omega$.

\begin{table}[t]
\begin{center}\small
\begin{tabular}{||c|ccc||}
\hline
Coupling& DDH `best' value & DDH range   & KMW\\
\hline
\rule{0pt}{3ex}
$h_\pi$ & $\phantom{-}4.6\cdot 10^{-7}$ & $ \phantom{-}(0.0\rightarrow11)\cdot10^{-7} $  &$\phantom{-}1.0\cdot 10^{-7}$\\
\rule{0pt}{2ex}
$h_\rho^0$ & $-11.4\cdot 10^{-7}$ &$\phantom{.}(-31\rightarrow 11)\cdot10^{-7}$ &$-1.9\cdot 10^{-7}$\\
\rule{0pt}{2ex}
$h_\rho^1$ & $-0.19\cdot 10^{-7}$ &$(-0.4\rightarrow 0.0)\cdot10^{-7}$& $-0.02\cdot 10^{-7}$\\
\rule{0pt}{2ex}
$h_\rho^2$ & $-9.5\cdot 10^{-7}$ &$\phantom{.}(-11\rightarrow -7.6)\cdot10^{-7}$& $-3.8\cdot 10^{-7}$\\
\rule{0pt}{2ex}
$ h_\rho^{1\,\prime}$ & $0$ &$0 $&$-2.2\cdot 10^{-7}$\\
\rule{0pt}{2ex}
$h_\omega^0$ & $-1.9\cdot 10^{-7}$ & $\phantom{.}(-10\rightarrow 5.7)\cdot10^{-7}$&$-1.1\cdot 10^{-7}$\\
\rule{0pt}{2ex}
$h_\omega^1$ & $-1.1\cdot 10^{-7}$ & $(-1.9\rightarrow -0.8)\cdot10^{-7}$&$-1.0\cdot 10^{-7}$\\
 \hline

\end{tabular}
\end{center}
\caption{\small Estimates of the DDH coupling constants. The first and second columns denote, respectively, the `best' values and reasonable range obtained in Ref.~\cite{Desplanques:1979hn} using $SU(6)$ symmetry arguments and the quark model. The third column denotes values obtained in Ref.~\cite{Kaiser1} using a non-linear chiral Lagrangian and a soliton description of the nucleon. The value of $h_\pi$ in the third column is from Ref.~\cite{Meissner1} which updated the result of Ref.~\cite{Kaiser1}.}
\label{table1}
\end{table}
The weak couplings appearing in the potential need to be fitted to data or, in absence of sufficient data, estimated in theoretical models. Several estimates exist in the literature and in Table \ref{table1} we give the two sets of estimates obtained in Refs.~\cite{Desplanques:1979hn, Kaiser1, Meissner1}. The first column denotes the `best' values obtained in Ref.~\cite{Desplanques:1979hn}  while the second column shows the reasonable range of these parameters \cite{Desplanques:1979hn}. The third column corresponds to values obtained in Refs.~\cite{Kaiser1, Meissner1}. The difference between the first and third columns reflects the significant uncertainty of these estimations.
Other sets of predictions can be found in, for example, Refs.~\cite{Dubovik:1986pj, Feldman:1991tj}.

\subsection{Comparing the potentials}\label{SecComp}
Most calculations in the literature have applied the DDH potential instead of the chiral EFT potential. It would therefore be useful to compare the parameters appearing in the different frameworks. It was already noted (see, for example, Ref.~\cite{Holsteinreview}) that at sufficiently low energy, the $\slashP$ potential is effectively saturated by $S\leftrightarrow P$ transitions. In this limit, the DDH potential collapses into a potential consisting of five independent contact interactions (one for each $S\leftrightarrow P$ transition). The resulting potential is identical to the pionless EFT potential given in Refs.~\cite{Girlanda:2008ts, Phillips:2008hn}.

If the energy is increased somewhat, the heavier mesons can still be integrated out, but the pion becomes dynamical. In this limit, the DDH potential consists of OPE supplemented by five effective contact interactions. The OPE in the DDH potential\footnote{It should be noted that if a regulator function is applied to the OPE part of the DDH potential then an additional term would appear on the right-hand side of the relation for $C_4$ given below. } is the same as the OPE in the LO EFT (see Eq.~\eqref{onepion}), while the five effective DDH contact interactions match onto the five interactions in Eq.~\eqref{CT}. By comparing the $S\leftrightarrow P$ transitions in the two frameworks, the following relations are obtained
\begin{eqnarray}\label{comparison}
\frac{C_0 + C_1}{ \Fp\Lambda_\chi^2}  &\sim&\frac{1}{m_N}\left[\frac{g_\omega h^0_\omega \chi_S}{m_\omega^2}c_{\omega}(0,\Lambda_{\omega}) - \frac{3 g_\rho h^0_\rho\chi_V}{m_\rho^2} c_{\rho}(0,\Lambda_{\rho})\right]\,\,\,,\nonumber\\
\frac{-C_0 + C_1}{ \Fp\Lambda_\chi^2} &\sim&\frac{1}{m_N}\left[ \frac{g_\omega h^0_\omega (2 +\chi_S)}{m_\omega^2}c_{\omega}(0,\Lambda_{\omega}) + \frac{g_\rho h^0_\rho (2 +\chi_V)}{m_\rho^2}c_{\rho}(0,\Lambda_{\rho}) \right]\,\,\,,\nonumber\\
\frac{C_2}{ \Fp\Lambda_\chi^2}  &\sim&\frac{1}{m_N}\left[ \frac{g_\omega h^1_\omega (2 +\chi_S)}{m_\omega^2}c_{\omega}(0,\Lambda_{\omega}) + \frac{g_\rho h^1_\rho (2 +\chi_V)}{m_\rho^2} c_{\rho}(0,\Lambda_{\rho})\right]\,\,\,,\nonumber\\
\frac{C_3}{ \Fp\Lambda_\chi^2} &\sim&-\frac{1}{m_N}\frac{g_\rho h^2_\rho (2 +\chi_V)}{\sqrt{6}\, m_\rho^2}c_{\rho}(0,\Lambda_{\rho})\,\,\,,\nonumber\\
\frac{C_4}{\Fp\Lambda_\chi^2} &\sim&\frac{1}{2m_N}\left[\frac{g_\omega h^1_\omega }{m_\omega^2}c_{\omega}(0,\Lambda_{\omega}) + \frac{ g_\rho( h^{1\,\prime}_\rho-h^1_\rho)}{m_\rho^2} c_{\rho}(0,\Lambda_{\rho})\right]\,\,\,.
\end{eqnarray}

\begin{table}[t]
\begin{center}\small
\begin{tabular}{||c|ccc||}
\hline
LEC& DDH `best' value &DDH range& KMW\\
\hline
\rule{0pt}{3ex}
$C_0$ & $\phantom{-}4.7\cdot 10^{-6}$ & $(-5.0\rightarrow 13)\cdot 10^{-6}$& $\phantom{-}0.89\cdot 10^{-6}$\\
\rule{0pt}{2ex}
$C_1$ & $\phantom{-}1.2\cdot 10^{-6}$ & $(-2.5\rightarrow 4.5)\cdot 10^{-6}$&$\phantom{-}0.11\cdot 10^{-6}$\\
\rule{0pt}{2ex}
$C_2$ & $-2.2\cdot 10^{-6}$ & $(-5.0\rightarrow -0.2)\cdot 10^{-6}$& $-0.66\cdot 10^{-6}$\\
\rule{0pt}{2ex}
$C_3$ & $\phantom{-}1.0\cdot 10^{-6}$ &$(0.8\rightarrow 1.2)\cdot 10^{-6}$& $\phantom{-}0.41\cdot 10^{-6}$\\
\rule{0pt}{2ex}
$ C_4$ & $\phantom{-}0.25\cdot 10^{-6}$ & $(-0.1\rightarrow 0.7)\cdot 10^{-6}$&$-0.049\cdot 10^{-6}$\\
 \hline

\end{tabular}
\end{center}
\caption{\small Predictions of the LECs $C_i$ using the resonance saturation relations in Eqs.~\eqref{comparison} and \eqref{TPEshift} and the estimates of the DDH couplings in Table \ref{table1}.}
\label{table2}
\end{table}

However, this is not the whole story. In the comparison of the EFT contact LECs with the DDH parameters, we have neglected the TPE contributions in Eq.~\eqref{NLOTPE}. As discussed in Refs.~\cite{Epelbaum:2001fm, Berengut:2013nh} these TPE contributions need to be taken into account for a sensible comparison with the one-meson exchange model.
To do so, we expand the TPE functions in powers of $q^2$, keeping the terms that connect $S$- and $P$-waves. Due to the TPE diagrams the relations connecting the DDH parameters and the contact terms are altered, with the LECs $C_2$ and $C_4$ in Eq.~\eqref{comparison} shifting into
\begin{eqnarray}\label{TPEshift}
\frac{C_2}{ \Fp\Lambda_\chi^2}\rightarrow \frac{C_2}{ \Fp\Lambda_\chi^2} +  \frac{ g_A^3 h_\pi}{2 \sqrt{2}\Fp}  \frac{8}{(4\pi \Fp)^2} \frac{\sqrt{\Lambda_S^2-4m_\pi^2}}{\Lambda_S}\,\,\,,\nonumber\\
\frac{C_4}{ \Fp\Lambda_\chi^2}\rightarrow \frac{C_4}{ \Fp\Lambda_\chi^2} -  \frac{ g_A h_\pi}{2 \sqrt{2}\Fp}  \frac{(2g_A^2-1)}{(4\pi \Fp)^2} \frac{\sqrt{\Lambda_S^2-4m_\pi^2}}{\Lambda_S}\,\,\,.
\end{eqnarray}

We are now in the position to compare the LECs appearing in both approaches. First of all, we can predict the values of the LECs $C_i$ using the estimates of the DDH coupling constants in Table \ref{table1}. These predicted values\footnote{To obtain these values of $C_i$ we have removed the cut-offs in the DDH potentials by setting $c_{\rho,\omega}=1$ in Eq.~\eqref{comparison}. Including the DDH cut-offs  with $\Lambda_\omega = 1.50$ GeV and $\Lambda_\rho =1.31$ GeV would reduce the predicted values by at most a factor $2$, depending on the $C_i$ under investigation. For the spectral cut-off appearing in Eq.~\eqref{TPEshift} we have used $\Lambda_S = 600$ MeV.} are given in Table \ref{table2}.  By dimensional analysis we expect roughly $C_i\sim 10^{-6}$ and most of the predicted LECs are indeed of this natural size. The main exception is $C_4$ which is smaller by an order of magnitude than $C_2$ in both sets of predictions, even though both are $\Delta I=1$ operators. We do not have a good explanation for this effect and it would be interesting to see if $C_4$ really takes such a small value in nature. However, the extraction of $C_4$ from data might be difficult because $C_4$ contributes to the same channels as the LO OPE potential which will, most likely, swamp the contributions from $C_4$. A possible way to extract $C_4$ would then be to measure an observable sensitive to OPE at several energies. $C_4$ can then be disentangled from $h_\pi$ by using the different energy dependence of the contributions. 

The $pp$ LAP has been calculated in both approaches \cite{deVries,Carlson:2001ma} and it is interesting to compare the analyses. The LAP is defined as the difference in cross section between an unpolarized target and a beam of positive and negative helicity, normalized to the sum of these cross sections. The $pp$ LAP has been measured for several beam energies ($13.6$, $45$, and $221$ MeV lab-energy \cite{Kistryn:1987tq, Eversheim:1991tg, Berdoz:2001nu}). In Ref. \cite{deVries} this observable was calculated using the N${}^3$LO chiral effective $P$-even potential~\cite{Epelbaum:2004fk} in combination with the NLO $\slashP$ potential (i.e. Eqs.~\eqref{CT} and \eqref{NLOTPE}). The LO $\slashP$ potential, and thus also $C_4$, does not contribute to the $pp$ LAP. Because only one combination of the contact interactions appears in $pp$ scattering, this particular combination was abbreviated as $C = - C_0 +C_1 +C_2 - C_3$ in Ref.~\cite{deVries}. The only other $\slashP$ LEC which appears up to NLO is the weak pion-nucleon coupling constant $h_\pi$. The two LECs were fitted to data in Ref.~\cite{deVries}. The same data was analyzed in Ref.~\cite{Carlson:2001ma} in order to fit the two combinations of DDH parameters which contribute to the $pp$ LAP. This analysis was slightly altered in Ref.~\cite{Holsteinreview} because the authors Ref.~\cite{Carlson:2001ma} uses CD-Bonn strong couplings instead of the strong parameters normally used in the DDH potential.

 \begin{figure}[t]
\centering
\includegraphics[scale=0.9]{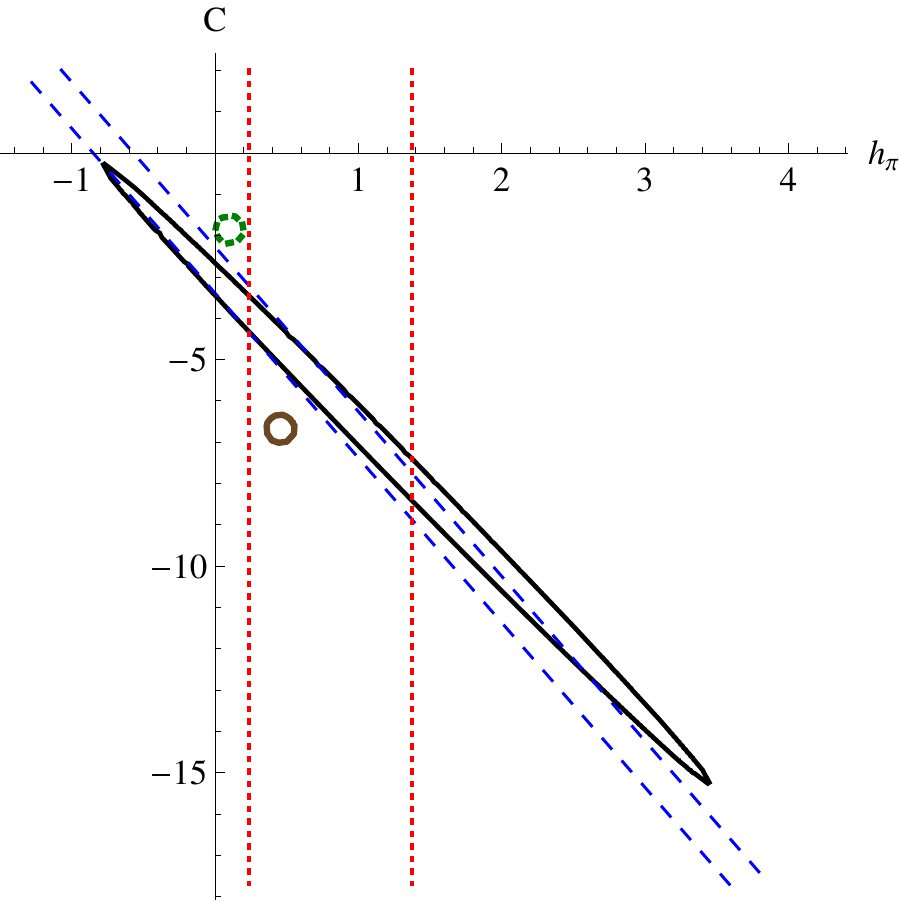}
\caption{Allowed ranges for the LECs $h_\pi$ and $C$ (both in units of $10^{-6}$) as obtained in different approaches from the data on the longitudinal analyzing power in $pp$ scattering. The black (solid) ellipse denotes a contour of total $\chi^2=2.71$ ($\chi^2_\mathrm{min}\simeq 0.75$ for $1$ d.o.f.) obtained in Ref.~\cite{deVries}. The blue (dashed) and red (dotted) lines are results obtained in the DDH model in Refs.~\cite{Carlson:2001ma,Holsteinreview} translated to values of $h_\pi$ and $C$ using resonance saturation. The brown (solid) and green (dotted) circles are, respectively, the DDH `best' value and the values obtained for the weak couplings in Refs. \cite{Kaiser1, Meissner1}.}
\label{DDH_EFT_1}
\end{figure}

At low energies the $pp$ LAP depends dominantly on the ${}^1S_0\leftrightarrow {}^3P_0$ transition. The above analysis tells us that the following combination of LECs should be compared
\begin{equation}\label{comparison2}
\frac{C}{ \Fp\Lambda_\chi^2} +  \frac{ g_A^3 h_\pi}{2 \sqrt{2}\Fp}  \frac{8}{(4\pi \Fp)^2} \frac{s}{\Lambda_S} \sim \frac{1}{ m_N}\left[ \frac{g_\omega  (2 +\chi_S)}{m_\omega^2} h^{pp}_\omega c_{\omega}(0,\Lambda_{\omega}) + \frac{g_\rho (2 +\chi_V)}{m_\rho^2}h^{pp}_\rho c_{\rho}(0,\Lambda_{\rho}) \right],
\end{equation}
with $h_\omega^{pp} = h_\omega^0 + h_\omega^1$ and $h_\rho^{pp} =h^0_\rho + h^1_\rho+ h^2_\rho/\sqrt{6}$. The spectral cut-off $\Lambda_S$ was varied\footnote{The results in Ref.~\cite{deVries} were cut-off independent to a large degree.} between $500$ and $700$ MeV in Ref.~\cite{deVries} while Ref.~\cite{Carlson:2001ma} used $\Lambda_\omega = 1.50$ GeV and $\Lambda_\rho =1.31$ GeV.

The allowed range for $C$ and $h_\pi$ is depicted by the black ellipse in Fig.~\ref{DDH_EFT_1} corresponding to a contour of a total $\chi^2=2.71$ and $\Lambda_S = 600$ MeV \cite{deVries}.  A very similar range was found in Ref.~\cite{Viviani}.
Refs.~\cite{Holsteinreview,Carlson:2001ma}  found that $h_\omega^{pp}$ and $h_\rho^{pp}$ are linearly correlated and, at $90\%$ confidence level, vary between
\begin{equation}\label{rangeDDH}
-66 \cdot 10^{-7}\leq h_\rho^{pp} \leq -18 \cdot 10^{-7}\,\,,\qquad +40 \cdot 10^{-7}\geq h_\omega^{pp} \geq -10 \cdot 10^{-7}\,\,\,.
\end{equation}
By using Eq.~\eqref{comparison2}, the allowed range for the DDH parameters can
be translated into a range for the LECs $C$ and $h_\pi$. This range is
depicted by the blue dashed lines in Fig.~\ref{DDH_EFT_1}. The analyses are in
good agreement as indicated by the significant overlap of the ellipse and the
band.  For comparison, we have used Eq.~\eqref{comparison2} and the
predictions of the weak DDH couplings in Table~\ref{table1} to obtain a
prediction for the EFT contact LEC $C$. The result is denoted by the brown
solid (DDH `best' values) and green dashed (KMW values) circles in Fig.~\ref{DDH_EFT_1}. Both predictions fall somewhat outside the allowed range, but it should be noted that the theoretical uncertainties of these model predictions are significant. In order to keep the plot clear we have not included the contour of the DDH reasonable range. This contour would roughly cover a range $-17\cdot 10^{-6}<C<3.5\cdot 10^{-6}$ for $0<h_\pi<1.1\cdot 10^{-6}$ and thus overlaps with the fitted values of the LECs.

So far we have only considered $S\leftrightarrow P$ transitions. In many observables (including the $pp$ LAP) higher-partial wave transitions play a role as well. The description of these transitions is quite different in the EFT and DDH potentials. For example, up to NLO in the EFT potential $P\leftrightarrow D$ transitions depend solely on $h_\pi$ via OPE and TPE diagrams. In the DDH framework OPE is also present, but in addition all terms in Eq.~\eqref{DDHpot} contribute to  $P\leftrightarrow D$ transitions as well. $P\leftrightarrow D$ transitions therefore depend on more DDH parameters than just $h_\pi$.
In $pp$ scattering, the only relevant transition is ${}^3 P_2\leftrightarrow {}^1D_2$ which means that the following combination of LECs should be compared
\begin{equation}\label{comparison3}
 -\frac{ g_A^3 h_\pi}{ \sqrt{2}\Fp}  \frac{1}{(4\pi \Fp)^2\mpi^2} \left(\frac{\Lambda_S^2-4m_\pi^2}{\Lambda_S^2}\right)^{3/2} \sim \frac{3}{ m_N}\left[ \frac{g_\omega  \chi_S}{m_\omega^4} h^{pp}_\omega c^\prime_\omega + \frac{g_\rho \chi_V}{m_\rho^4}h^{pp}_\rho c^\prime_\rho \right]\,\,\,,
\end{equation}
in terms of the functions
\begin{equation}\label{cprime}
c^\prime_{\rho,\omega} = 1-\frac{3m_{\rho,\omega}^4}{\Lambda_{\rho,\omega}^4}+\frac{2m_{\rho,\omega}^6}{\Lambda_{\rho,\omega}^6}\,\,\,.
\end{equation}
Relations for other $P\leftrightarrow D$ transitions can be found in Appendix~\ref{AppA}.

An allowed range for $h_\pi$ can now be extracted by inserting Eq.~\eqref{rangeDDH} into Eq.~\eqref{comparison3}. This range is indicated by the red dotted lines in Fig.~\ref{DDH_EFT_1}. The central values for $h_\pi$ agree well between the methods, but the range allowed in the EFT framework is larger. The difference might be related to the so-called `crossing points' discussed in Ref.~\cite{deVries}, which limited the discriminating power of the fits with respect to $h_\pi$. Nevertheless, we conclude that the analyses of the $pp$ data in Refs.~\cite{Carlson:2001ma,Holsteinreview} in the DDH model and Refs.~\cite{deVries, Viviani} using chiral potentials are in good agreement.

\section{The next-to-next-to-leading-order parity-violating potential }\label{N2LOpot}
The chiral NLO $\slashP$ potential depends on six LECs consisting of the weak pion-nucleon coupling $h_\pi$ and five contact LECs $C_i$. The five possible $S\leftrightarrow P$ transitions depend on all six LECs, but, because of the short-range nature of the contact interactions, higher partial-wave transitions depend solely on $h_\pi$. Theoretical calculations show that $h_\pi$ might be very small \cite{Kaiser1, Meissner1, hpilatt}, something which is also indicated by the data on ${}^{18}F$ gamma-ray emission \cite{Adelberger:1983zz,Adelberger:1983zz2}
\begin{equation}
|h_\pi| < 1.3 \cdot 10^{-7}\,\,\,.
\end{equation}
This possible smallness implies that higher $\slashP$ partial-wave transitions\footnote{Additional contribution to $S\leftrightarrow P$ transitions are most likely swamped by the contact LECs $C_i$ which are not expected to be small.} might obtain important, or even dominant, contributions from higher-order corrections involving additional LECs. It might therefore be necessary to include these higher-order corrections in the analysis of future experimental results. Here we extend the calculation of the $\slashP$ $N\!N$ potential up to N${}^2$LO and study the corrections appearing at this order. We start by listing the new operators that need to be taken into account.

\subsection{The next-to-next-to-leading-order Lagrangian}
If we increase the chiral index by one, several new interactions start playing a role. We present these interactions here and afterwards discuss how they contribute to the $\slashP$ potential. First of all, we require the well-known $P$-even corrections to the $\pi\pi N$-vertex \cite{Bernard:1995dp}
\begin{eqnarray}\label{pipiN}
\mathcal L_{\pi\pi N} &=&\mpi^2\left(4 c_1  - \frac{2 c_1 \vec \pi^{\,2}}{\Fp^2}\right)\Nb N + \frac{c_2}{\Fp^2}(v\cdot  \partial \vec \pi)^2 \Nb N + \frac{c_3}{\Fp^2}(  \partial_\mu \vec \pi)^2 \Nb N\nonumber\\
&& + \frac{ c_4}{\Fp^2} \epsilon^{\al\bt\mu\nu}v_\al \epsilon^{abc} ( \partial_\mu \pi^a)( \partial_\nu \pi^b) \Nb S_\beta \tau^c N+\dots\,\,\, ,
\end{eqnarray}
in terms of the completely antisymmetric tensor $\epsilon^{\alpha\beta\mu\nu}$ $(\epsilon^{0123}=+1)$ and the nucleon velocity $v^\mu = (1,\vec 0)^t$ and spin $S^\mu = (0,\vec \sigma/2)^t$ in the nucleon rest frame. Here and in the following equations the dots denote operators containing more pions which are related by chiral symmetry. Such terms only contribute to the potential at higher order than considered here and therefore we do not write them explicitly.

Next we list the required additional $\slashP$ interactions. A complete set of operators was recently constructed in Ref.~\cite{Viviani}, but most operators do not contribute to the N${}^2$LO potential, and here we only require a much smaller set.
 Three pion-nucleon vertices appear with one derivative \cite{Kaplan:1992vj}
\begin{equation}\label{hv}
\mathcal L_{\slashPsub } = h^v_0 (v\cdot \partial \vec \pi)\cdot \Nb \vec\tau N+ h^v_1  (v\cdot \partial \pi^3) \Nb N+ h^v_2  (v\cdot \partial \pi^3) \Nb \tau^3 N +\cdots\,\,\,.
\end{equation}
With two derivatives, a few more interactions can be written. First we consider recoil corrections to the vertices in Eq. \eqref{hv}
\begin{eqnarray}\label{hvrecoil}
\mathcal L_{\slashPsub } &=& \frac{h^v_0}{2 m_N} (\partial_\mu \vec \pi)\cdot \Nb i(\partial_\mu - \partial_\mu^\dagger)\vec\tau N  +\frac{h^v_1}{2 m_N} (\partial_\mu \pi^3) \Nb i(\partial_\mu - \partial_\mu^\dagger)N \nonumber\\
&& +\frac{h^v_2}{2 m_N} (\partial_\mu \pi^3)\Nb i(\partial_\mu - \partial_\mu^\dagger)\tau^3 N \,\,\,.
\end{eqnarray}
Two terms appear with new LECs
\begin{equation}\label{2deriv}
\mathcal L_{\slashPsub } = \frac{h^{(2)}_\pi}{\sqrt{2}} \Nb (\partial^2 \vec \pi\times \vec \tau)^3 N+  \frac{h_{m} \mpi^2}{\sqrt{2}} \Nb ( \vec \pi\times \vec \tau)^3 N\,\,\,,
\end{equation}
where the second term emerges due to an insertion of the quark mass.
In addition, we require the following two $\slashP$ $\pi\pi NN$ vertices  \cite{Kaplan:1992vj} containing one derivative
\begin{equation}\label{hpipi}
\mathcal{L}_{\slashPsub }=\frac{h^{\pi \pi}_1}{\Fp}(\vec \pi \times \partial \vec \pi)^3 \Nb S^\mu N+\frac{h^{\pi \pi}_2}{F_{\pi}}\left[\partial_{\mu}\pi^3\Nb(
\vec \pi \times \vec \tau)^3 S^{\mu} N+(\vec \pi\times \partial_{\mu}\vec \pi)^3 \Nb S^{\mu} \tau^3 N\right].
\end{equation}
Finally, at this order the first $\slashP$ three-pion vertex appears \cite{Viviani}
\begin{equation}\label{3pion}
\mathcal L_{\slashPsub} = \Delta_{\pi} (\vec \pi \times \partial_\mu \vec \pi)^3\partial^\mu \pi^3\,\,\,.
\end{equation}

Additional $\slashP$ $N\!N$ contact interactions do not appear. Such terms would induce $P\leftrightarrow D$-transitions and require two more derivatives than the interactions in Eq. \eqref{CT}. Consequently, they first appear at N${}^3$LO and can be neglected.

 We conclude that, with respect to the NLO $\slashP$ potential, the N${}^2$LO potential depends at most on the eight additional LECs appearing in Eqs.~\eqref{hv}-\eqref{3pion}. To get an idea of the sizes of the new LECs we apply naive dimensional analysis to obtain
 \begin{equation}\label{NDA2}
 h^v_i \sim \Lambda_\chi\,h^{(2)}_\pi\sim \Lambda_\chi\,h_m \sim h^{\pi\pi}_i \sim \Delta_\pi  \sim \frac{h_\pi}{\Lambda_\chi}\sim \frac{C_i}{\Lambda_\chi}\sim \mathcal O(G_F F_\pi)\,\,\,.
 \end{equation}
 We now turn to the actual calculation of the N${}^2$LO potential starting with corrections to the OPE potential.

 \subsection{One-pion-exchange corrections}\label{OPEcor}
At higher orders in the chiral expansion, the OPE potential in Eq.~\eqref{onepion} might need to be extended or adjusted. Let us first consider OPE diagrams involving the subleading $\slashP$ $\pi N$ vertices in Eq.~\eqref{hv} and the LO $P$-even $\pi N$ vertex. The sum of these diagrams gives rise to the following potential
\begin{eqnarray}\label{hvOPE}
V(q) &=& -\frac{g_A}{2\Fp}\bigg \{\left[ \left(h^v_0 + \frac{1}{3} h^v_2\right)\vec \tau_1 \cdot \vec \tau_2 + h^v_1 \frac{(\vec\tau_1 + \vec \tau_2)^3}{2}+\frac{h^v_2}{3} (3\tau_1^3 \tau_2^3-\vec \tau_1\cdot \vec \tau_2)\right]\nonumber\\
&&\times(\vec \sigma_1 + \vec \sigma_2)\cdot \vec q  + h^v_1 \frac{(\vec\tau_1 - \vec \tau_2)^3}{2} (\vec \sigma_1 - \vec \sigma_2)\cdot \vec q\bigg\} \frac{q^0}{ \mpi^2+q^2}\,\,\,,
\end{eqnarray}
where $q^0$ is the energy of the exchanged pion. Since\footnote{Here, we follow Ref.~\cite{Epelbaum:2004fk} and count an inverse power of $m_N$ as two inverse powers of $\Lambda_\chi$, that is $1/m_N \sim k/\Lambda_\chi^2$ where $k$ is the typical momentum scale in the process. } $q^0 \sim q^2/m_N \sim \Or (q^3/\Lambda_\chi^2)$ and $h_\pi/h^v_i \sim \Or(1/\Lambda_\chi)$ the above terms appear at N${}^2$LO. However, it is not hard to see that all these terms induce transitions in which the total spin and total isospin of the nucleon pair either both stay the same or both change. Since a $\slashP$ transition simultaneously requires a change of the angular momentum by one unit, the Pauli principle ensures that Eq.~\eqref{hvOPE} vanishes when acting between two nucleons. In what follows, we only list the non-vanishing contributions.

Let us now consider the vertices in Eq.~\eqref{hvrecoil}. They lead to the following OPE potential
\begin{eqnarray}\label{hvrecoilOPE}
V(q) &=& -\frac{g_A}{4 \Fp m_N}\bigg \{\left[ \left(h^v_0 + \frac{1}{3} h^v_2\right)\vec \tau_1 \cdot \vec \tau_2 + h^v_1 \frac{(\vec\tau_1 + \vec \tau_2)^3}{2}+\frac{h^v_2}{3}(3\tau_1^3 \tau_2^3-\vec \tau_1\cdot \vec \tau_2)\right]\nonumber\\
&&\times(\vec \sigma_1 - \vec \sigma_2)\cdot \vec q  + h^v_1 \frac{(\vec\tau_1 - \vec \tau_2)^3}{2} (\vec \sigma_1 + \vec \sigma_2)\cdot \vec q\bigg\} \frac{\vec p^{\,2}-\vec p^{\,\prime 2}}{ \mpi^2+q^2}\,\,\,,
\end{eqnarray}
which appears at N${}^2$LO and does not automatically vanish.

Next we consider the operators in Eq.~\eqref{2deriv}, which add the following terms
\begin{eqnarray}
V&=& - \frac{g_{A}}{ 2\sqrt{2} F_\pi} i(\vec \tau_1\times \vec \tau_2)^3\left(q^2 h_\pi^{(2)} + \mpi^2 h_m \right) \frac{(\vec \sigma_1+\vec \sigma_2)\cdot \vec q }{\mpi^2+q^2}\nonumber\\
&=&- \frac{g_{A}}{ 2\sqrt{2} F_\pi} i(\vec \tau_1\times \vec \tau_2)^3\left(h_\pi^{(2)} + \frac{\mpi^2(h_m - h_\pi^{(2)})}{\mpi^2 +q^2}  \right) (\vec \sigma_1+\vec \sigma_2)\cdot \vec q\,\,\,.
\label{onepion2deriv}
\end{eqnarray}
A comparison with Eqs.~\eqref{onepion} and \eqref{CT} shows that the above terms can be absorbed into $h_\pi$ and $C_4$
\begin{equation}\label{absorb}
h_\pi \rightarrow h_\pi + \mpi^2 (h_m - h_\pi^{(2)})\,\,, \qquad \frac{C_4}{2\Fp \Lambda_\chi^2}  \rightarrow  \frac{C_4}{2\Fp \Lambda_\chi^2}-\frac{g_A h_\pi^{(2)}}{2\sqrt 2 \Fp}\,\,\,.
\end{equation}
As a consequence the LECs in Eq.~\eqref{2deriv} do not appear independently in the N${}^2$LO $\slashP$ potential.

\begin{figure}[t]
\centering
\includegraphics[scale=0.8]{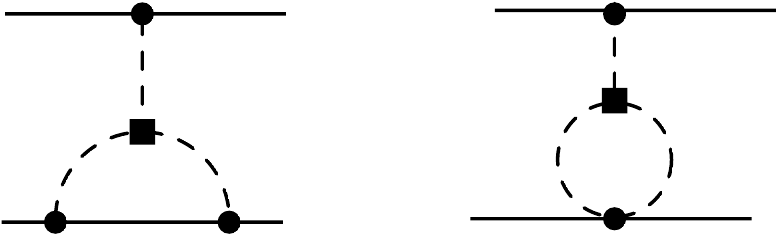}
\caption{Diagrams contributing to the $\slashP$ $N\!N$ potential at N$^2$LO and containing the three-pion vertex (denoted by the square) in Eq.~\eqref{3pion}. Solid and dashed lines denote, respectively, nucleon and pion propagators while circles represent LO $P$-even vertices.}
\label{3pionpot}
\end{figure}

Apart from the above corrections to the potential, we need to consider one-loop corrections to the $\slashP$ pion-nucleon interaction. Most diagrams have been calculated up to N${}^2$LO in Refs.~\cite{Zhu:2000fc, Viviani} where it was concluded that all corrections can be absorbed into the LO coupling $h_\pi$. We have checked this calculation and found the same conclusion. The three-pion vertices in Eq.~\eqref{3pion} were not considered before and could lead to a N${}^2$LO contribution to the $\slashP$ potential via the diagrams in Fig.~\ref{3pionpot}.  However, these diagrams vanish such that $\Delta_\pi$ plays no role in the N${}^2$LO potential.

So far, we have considered corrections to the $\slashP$ $\pi N$ vertex. In addition, we must consider OPE diagrams which include corrections to the $P$-even $\pi N$ vertex in combination with $h_\pi$. The subleading $\pi N$ vertex proportional to $g_A/m_N$ gives a correction to the $\slashP$ potential proportional to $h_\pi (v\cdot q/m_N) \sim h_\pi (q^2/m_N^2) \sim h_\pi (q^4/\Lambda_\chi^4)$ and thus appears at N${}^3$LO. Corrections proportional to the quark masses can be absorbed into $g_A$ (which we have already done by using $g_A =1.29$), while operators with additional derivatives acting on the pion can, as in Eq.~\eqref{absorb}, be absorbed into $h_\pi$ and $C_4$.

Finally, we consider isospin-breaking corrections. The most important corrections are due to the pion-mass splitting. Because in the LO potential only the exchange of charged pions contributes, we already take these corrections into account by using $\mpi = 139.57$ MeV. In TPE and higher-order OPE diagrams the neutral pion also propagates, but the error made in using the charged-pion mass is of higher order than considered here. The same holds for the effects of the nucleon-mass splitting in TPE diagrams~\cite{Epelbaum:2004fk}.  In principle isospin-breaking $P$-even $\pi N$ couplings (see Ref.~\cite{Fettes:1998ud}) in combination with $h_\pi$ would appear in the OPE potential at N${}^2$LO. However, such diagrams vanish because the isospin-breaking $P$-even vertices couple only neutral pions while the LO $\slashP$ vertex only couples charged pions.

We conclude that, apart from renormalization of LO LECs, the only corrections to the OPE potential are those in Eq.~\eqref{hvrecoilOPE}. Naively one might think that this potential should vanish in nucleon-nucleon scattering since it is proportional to $\vec p^{\,2} - \vec p^{\,\prime 2}$ which is zero on-shell. However, the potential needs to be inserted into the Lippmann-Schwinger equation whose solution also depends on the off-shell potential. Nevertheless, the contributions from Eq.~\eqref{hvrecoilOPE} turn out to be small compared to the N${}^2$LO TPE contributions calculated in the next section. We discuss this in more detail in Sect.~\ref{remainder}.

\subsection{Two-pion-exchange corrections}
\begin{figure}[t]
\centering
\includegraphics[scale=0.8]{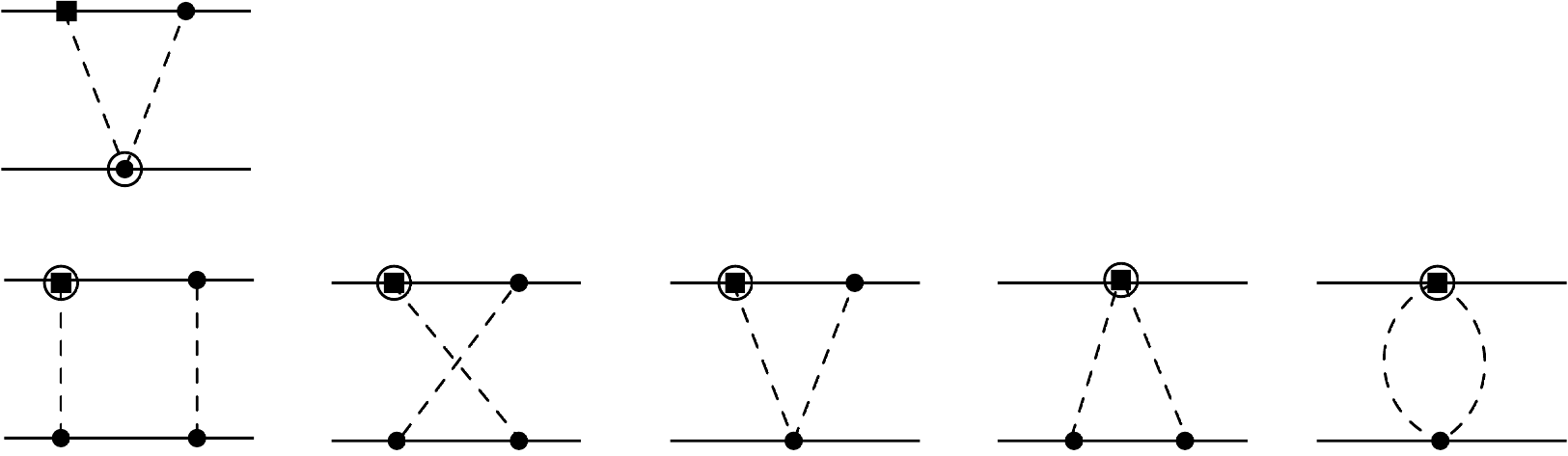}
\caption{Diagrams contributing to the $\slashP$ $N\!N$ potential at N$^2$LO. The squares denote LO $\slashP$ vertices, while circled circles and squares denote subleading vertices. The other notation is as in Fig.~\ref{3pionpot}.}
\label{TPE_N2LO}
\end{figure}

At N${}^2$LO we need to take into account TPE diagrams which, apart from LO vertices, include one of the subleading interactions in Eqs.~\eqref{pipiN}, \eqref{hv}, or \eqref{hpipi}. The interactions in Eq.~\eqref{hvrecoil} and \eqref{2deriv} only appear at higher orders in TPE diagrams.  The possible topologies of the N${}^2$LO TPE diagrams are shown in Fig. \ref{TPE_N2LO} where the circled circles and squares denote, respectively, subleading $P$-even and -odd vertices.

The diagram in the first row of Fig.~\ref{TPE_N2LO} contains one of the subleading $P$-even $\pi\pi N$ interactions in Eq. \eqref{pipiN}. In fact, most of these diagrams vanish and only an insertion of $c_4$ gives a non-vanishing result
\begin{eqnarray}\label{c4}
V (q, \Lambda_S) =   \frac{g_{A} h_\pi }{ 2\sqrt{2} F_\pi} \frac{ \pi c_4}{
  (4\pi \Fp)^2}  (\vec \tau_1+ \vec \tau_2)^3 i(\vec \sigma_1\times \vec \sigma_2)\cdot \vec q\, F(q,\Lambda_S)\,\,\,,
\end{eqnarray}
in terms of the spectrally regularized loop function
\begin{equation}
F(q,\Lambda_S) = \frac{\omega^2}{q} \arctan \frac{q\Lambda_S-2q\mpi}{q^2 +2\mpi \Lambda_S}\,\,\,.
\end{equation}
The result in Eq. \eqref{c4} is enhanced by a factor $\pi$ over the power-counting expectation. Combined with the fact that $c_4 = 3.4$ GeV$^{-1}$~\cite{Epelbaum:2004fk, Buettiker:1999ap} is somewhat larger than might be expected (due to underlying $\Delta$- and $\rho$-resonance contributions
  \cite{Bernard:1996gq}), the contribution in Eq. \eqref{c4} could be significant. We investigate this in more detail in the next section, where we investigate its effects on $\slashP$ $pp$ scattering.

Next we consider the diagrams in the second row of Fig. \ref{TPE_N2LO} which are due to an insertion of the subleading $\slashP$ $\pi NN$ or $\pi\pi NN$ vertices in Eqs. \eqref{hv} and \eqref{hpipi}. It turns out the total result can be written in rather compact form
\begin{eqnarray}\label{TPE_total}
V (q, \Lambda_S) &=& \frac{g_{A}  }{ 4 F_\pi} \frac{ \pi }{ (4\pi \Fp)^2} \bigg\{\left[ \bar h_0 +(\vec \tau_1\cdot \vec\tau_2) \bar h_0^\prime
 + \frac{(\vec \tau_1+\vec \tau_2)^3}{2} \bar h_1 +\frac{\vec \tau_1\cdot \vec \tau_2-3 \tau_1^3 \tau_2^3 }{2} \bar h_2\right]\nonumber\\
 &&\times  i (\vec \sigma_1 \times \vec \sigma_2)\cdot \vec q  +  \bar h_1^\prime  \,
i (\vec \tau_1\times \vec \tau_2)^3 (\vec \sigma_1+ \vec \sigma_2)\cdot \vec q
 \,\left(1-\frac{2\mpi^2}{\omega^2}\right)\bigg\}F(q,\Lambda_S)\,\,\,,
\end{eqnarray}
where we redefined the LECs
\begin{eqnarray}
\bar h_0 &=& \frac{g_A^2}{2} (3h_0^v +  h_2^v)\,\,\,,\qquad
\bar h_0^\prime = g_A^2 ( h_0^v + \frac{1}{3}  h_2^v)\,\,\,,\nonumber\\
\bar h_1 &=&  \frac{g_A^2}{2} h^v_1 - g_A h_1^{\pi\pi}\,\,\,,\qquad
\bar h_2 = \frac{g_A^2}{3}h^v_2+ g_Ah^{\pi\pi}_2\,\,\,,\nonumber\\
\bar h_1^\prime &=& -\frac{g_A^2}{2}h^v_1\,\,\,.
\end{eqnarray}
Actually, only the second and fourth diagram in the second row of 
Fig. \ref{TPE_N2LO} give nonvanishing contributions as can be inferred from the 
scaling with $g_A^2$ and $g_A^3$ of the complete expression. 
We conclude that the N${}^2$LO TPE potential depends on an additional five combinations of LECs.

\subsection{A refit of {\boldmath$h_\pi$} using part of the N{\boldmath$^2$}LO potential} \label{refitc4}
In order to get a feeling for the importance of the N$^2$LO TPE terms obtained in the previous section, we investigate the role played by the correction in Eq. \eqref{c4}. The reasons for choosing this correction are threefold. First, this correction involves no new unknown LECs with respect to the NLO potential. Secondly, because $c_4=3.4$ GeV$^{-1}$ is somewhat larger than expected from dimensional analysis, this term might actually dominate the N$^2$LO potential. Of course, this last point needs to be verified by a complete extraction of all relevant LECs. Thirdly, the contribution in Eq. \eqref{c4} contributes to $P$ violation in $pp$ scattering, a process which has recently been studied in detail in Ref. \cite{deVries}. It is straightforward to extend the formalism built there to include higher-order corrections to the TPE potential.

In Ref.~\cite{deVries} the LECs $h_\pi$ and $C$ (see Sec.~\ref{SecComp} for the definition of $C$) were fitted to the data of the $pp$ LAP. Because only three data points exist, the uncertainty in the fits was substantial.  It was concluded that the following ranges for the LECs are allowed at the level of a total $\chi^2 =2.71$ (see Fig.~\ref{DDH_EFT_1})
 \begin{eqnarray}\label{fitNLO}
h_\pi &=& (1.1\pm 2 )\cdot 10^{-6}~,\nonumber\\
C &=& (-6.5\pm 8)\cdot 10^{-6}~,
\end{eqnarray}
with the LECs strongly correlated. The error is dominated by the lack of data points and the significant experimental uncertainties in the existing points. The theoretical uncertainty, estimated by varying the cut-off parameters, only plays a minor role. Similar conclusions were found in Ref.~\cite{Viviani}.

\begin{figure}[t]
\centering
\includegraphics[width=0.49\textwidth]{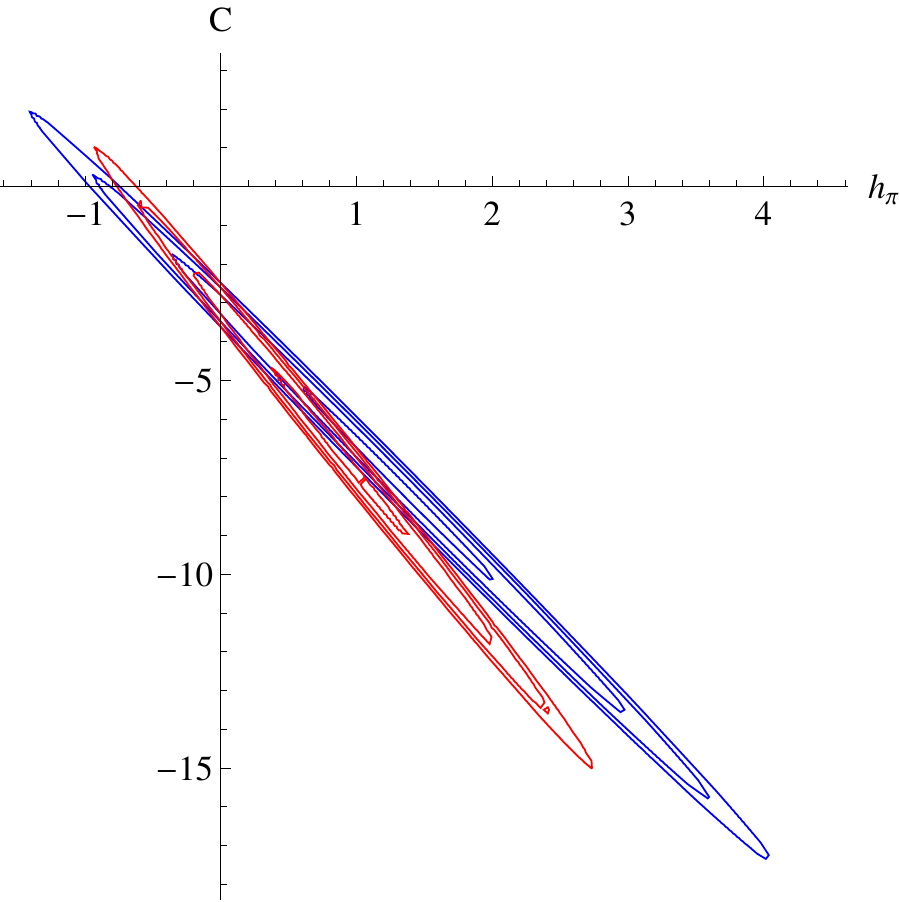}
\includegraphics[width=0.49\textwidth]{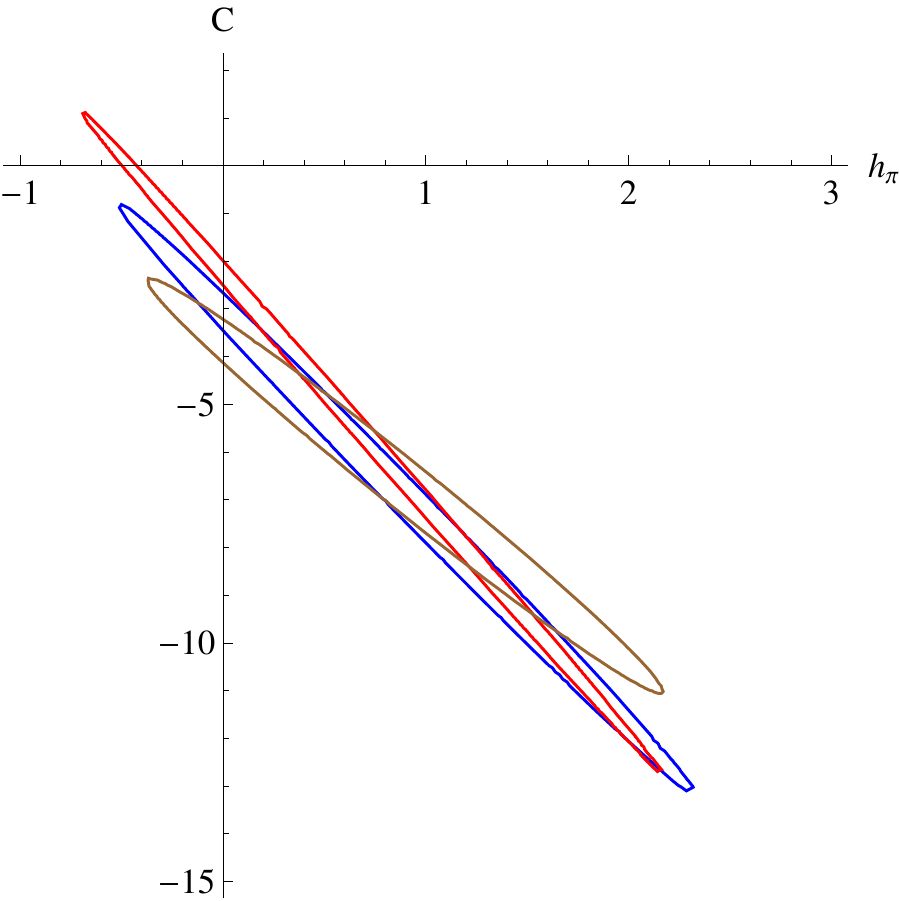}
\caption{Contours of constant $\chi^2$ in the $h_\pi-C$ plane (both in units
of $10^{-6}$). The left plot shows contours of total $\chi^2=1,2,3,4$   for
the NLO potential (blue contours) ($\chi^2_\mathrm{min}\simeq 0.75$ for $1$ d.o.f.) and the extended NLO potential (red contours) ($\chi^2_\mathrm{min}\simeq 0.7$ for $1$ d.o.f.), while the right plot shows contours
of total $\chi^2=2.71$ ($\chi^2_\mathrm{min}$ between $0.5$ and $1.2$ for $1$ d.o.f. depending on the cut-off combination) for three different cut-off combinations and the extended NLO potential.  }
 \label{Contour}
\end{figure}

We now repeat the analysis of Ref.~\cite{deVries} (and refer the reader there
for all details), but add to the potential the N$^2$LO correction in
Eq.~\eqref{c4}. Since this term involves no new LECs, the extraction of the
LECs works in exactly the same way. We find that, with the additional
correction, the fit is improved by a small margin, but this does not provide
too much information because the NLO fit was already very good. More
interesting is how the additional correction affects the extraction of the
LECs. In the left panel of Fig.~\ref{Contour}, we plot contours of constant
total $\chi^2 =1,2,3,4$ using one particular cut-off combination (the
spectral function cut-off was chosen as $\Lambda_S = 600$ MeV while the cut-off appearing in the Lippmann-Schwinger equation was taken as $\Lambda_{LS} = 550$ MeV, see Ref.~\cite{deVries}). The blue and red contours are associated with the NLO potential and the NLO potential supplemented by Eq.~\eqref{c4}, respectively. The N$^2$LO correction does not affect the values of the LECs by a large amount and the contours mostly overlap. This indicates that the power counting is working well (although it must be stressed that we did not analyze the full N$^2$LO potential).

In the right panel of Fig.~\ref{Contour} we plot contours of a constant $\chi^2 =2.71$ for three cut-off combinations~\cite{deVries}. From this plot we obtain the following allowed ranges for the LECs
 \begin{eqnarray}\label{fitNLOplus}
h_\pi &=& (0.8\pm 1.5 )\cdot 10^{-6}\,\,\,,\nonumber\\
C &=& (-5.5\pm 7)\cdot 10^{-6}\,\,\,.
\end{eqnarray}
The new fit values are approximately $20\%$ smaller than Eq. \eqref{fitNLO}, but fall within the error margins.  It should be mentioned again that the $pp$ LAP does not depend on the LO $\slashP$ potential in Eq.~\eqref{onepion}. This implies that the effects of the N${}^2$LO correction on observables where the LO potential does contribute will be smaller than $20\%$.

\subsection{The remaining corrections}\label{remainder}
\begin{figure}[t]
\centering
\includegraphics[width=0.5\textwidth]{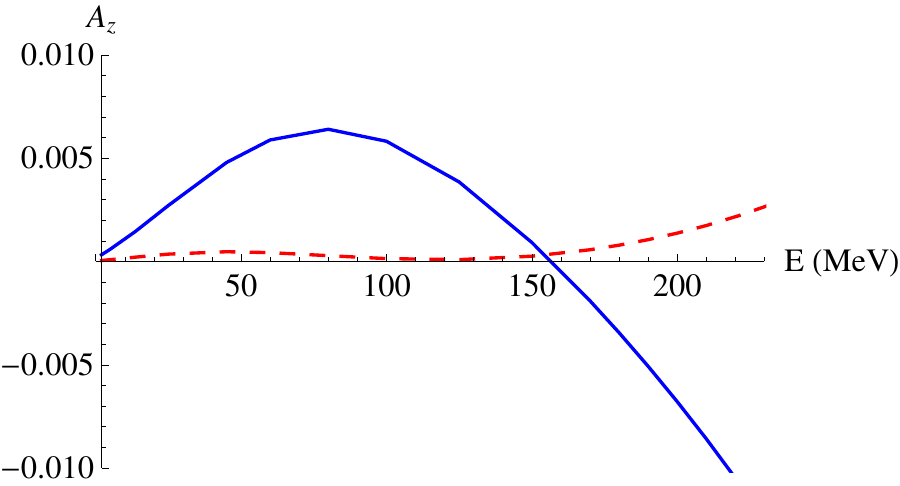}
\caption{Contributions to the integrated analyzing power in $pp$ scattering (angular range $23^\circ-52^\circ$ see Ref.~\cite{deVries}) in units of $\Lambda_\chi\,\bar h^{pp}$ as a  function of lab energy. The red (dashed) and blue (solid) lines denote, respectively, N${}^2$LO OPE and TPE contributions from Eq.~\eqref{remainderN2LO}. }
 \label{Az_N2LO_OTPE}
\end{figure}

\begin{figure}[t]
\centering
\includegraphics[width=0.49\textwidth]{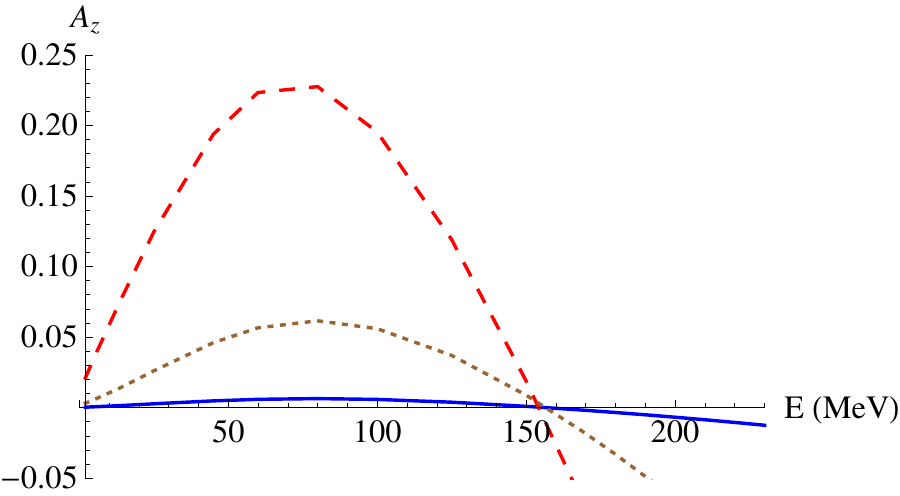}
\includegraphics[width=0.49\textwidth]{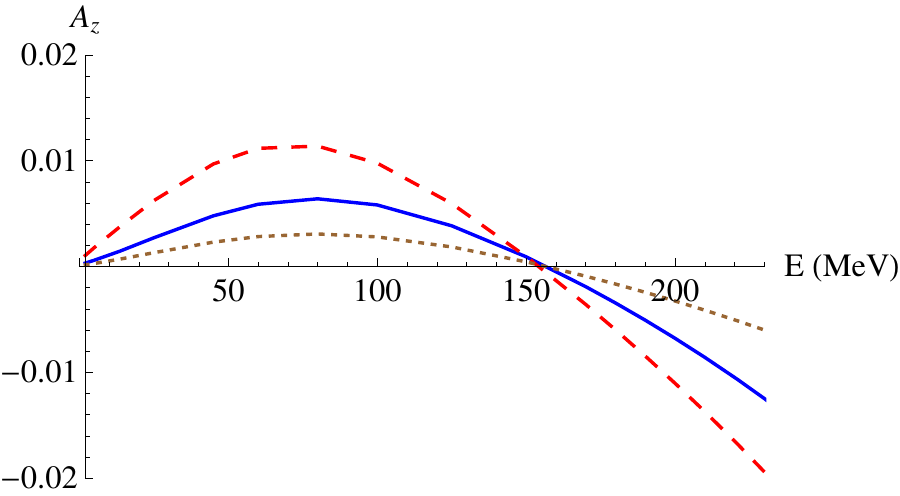}
\caption{Similar as Fig.~\ref{Az_N2LO_OTPE}. The blue (solid) line denotes N${}^2$LO TPE contributions from Eq.~\eqref{remainderN2LO}, while the red (dashed) and brown (dotted) lines denote contributions from, respectively, the NLO and N${}^2$LO TPE terms in Eqs.~\eqref{NLOTPE} and \eqref{c4}. The right panel is the same but the red (dashed) and brown (dotted) line are suppressed by a factor $20$, see text. }
 \label{Az_N2LO_TPE}
\end{figure}

The next step is the discussion of the remaining N$^2$LO corrections.  These consist of the OPE terms in Eq.~\eqref{hvrecoilOPE} and the TPE terms in Eq.~\eqref{TPE_total}. We study them by calculating their contributions to the $pp$ LAP. In the case of $pp$ scattering, the  N$^2$LO potential depends on two combinations of LECs
\begin{eqnarray}\label{remainderN2LO}
V_{\mathrm{OPE}} &=&\bar h^{pp}_{\mathrm{OPE}}\left(\frac{g_A}{4 \Fp m_N}\right)(\vec \sigma_1 - \vec \sigma_2)\cdot \vec q \,\,\frac{\vec p^{\,2}-\vec p^{\,\prime 2}}{\mpi^2+\vec q^2}\,\,\,,\nonumber\\
V_{\mathrm{TPE}} &=& \bar h^{pp}_{\mathrm{TPE}}\left( \frac{g_{A}  }{ 4 F_\pi} \frac{ \pi }{(4\pi\Fp)^2}\right)i (\vec \sigma_1 \times \vec \sigma_2)\cdot \vec q\, F(q,\Lambda_S)\,\,\,,
\end{eqnarray}
in terms of the couplings
\begin{equation}\label{defhbar}
\bar h^{pp}_{\mathrm{OPE}} = -(h^v_0+h^v_1+h^v_2)\,\,,\qquad \bar h^{pp}_{\mathrm{TPE}} =\frac{g_A}{2}\left[ g_A ( 5h^v_0+h^v_1+h^v_2)-2(h_1^{\pi\pi}+h_2^{\pi\pi})\right]\,\,\,.
\end{equation}
In principle these two couplings are independent, but they are expected to be of the same order of magnitude. In order to study the relative sizes of the two terms in Eq.~\eqref{remainderN2LO} we therefore assume $\bar h^{pp}_{\mathrm{OPE}} = \bar h^{pp}_{\mathrm{TPE}} \equiv \bar h^{pp}$. In Fig.~\ref{Az_N2LO_OTPE} we show the contributions to the integrated $pp$ LAP as a function of lab energy. It is clear that the OPE corrections are significantly smaller than the TPE corrections over the whole relevant energy range (the $pp$ LAP has been measured at $13.6$, $45$, and $221$ MeV) apart from a small region around $150$ MeV where both contributions are very small. The suppression of the OPE corrections might be due to the dependence on $\vec p^{\,2}- \vec p^{\,\prime\,2}$ which vanishes on-shell. We conclude that to good approximation we can neglect the N${}^2$LO OPE corrections and only consider the N${}^2$LO TPE terms.

We then need to investigate the size of the TPE corrections in Eq.~\eqref{remainderN2LO} with respect to the NLO TPE potential in Eq.~\eqref{NLOTPE} and its correction in Eq.~\eqref{c4}. The latter two depend on $\bar h_\pi$ which is expected to scale roughly as $h_\pi \sim \bar \Lambda_\chi\,h^{pp}_{\mathrm{TPE}}$ (see Eq.~\eqref{NDA2}). In the left panel of Fig.~\ref{Az_N2LO_TPE} we show the contributions to the $pp$ LAP from the three different terms using exactly this scaling. From the plot it becomes clear that if $h_\pi$ and $\bar h^{pp}_{\mathrm{TPE}}$ are naturally sized, the NLO potential completely dominates while the N${}^2$LO term proportional to $h_\pi c_4$ gives a $20\%$ correction (in agreement with the results in Sec.~\ref{refitc4}). The remaining N${}^2$LO correction is smaller by a factor \mbox{$2\sqrt{2} c_4 \Lambda_\chi \simeq 10$} than the $h_\pi c_4$ correction and therefore negligible.

This would be the scenario if all LECs are of the expected size. However, theory and experiments indicate that $h_\pi$ might actually be an order of magnitude smaller than expected from dimensional analysis. In the right panel of Fig.~\ref{Az_N2LO_TPE} we therefore show the results if we use $h_\pi = \bar h^{pp}_{\mathrm{TPE}} \times (1 \,\mathrm{GeV})/20$. In this case, the TPE corrections from Eq.~\eqref{remainderN2LO} are of similar size as the NLO TPE corrections. This indicates that the N${}^2$LO corrections can become relevant if it turns out that $|h_\pi|  \lesssim 10^{-7}$.

\begin{figure}[t]
\centering
\includegraphics[scale=0.8]{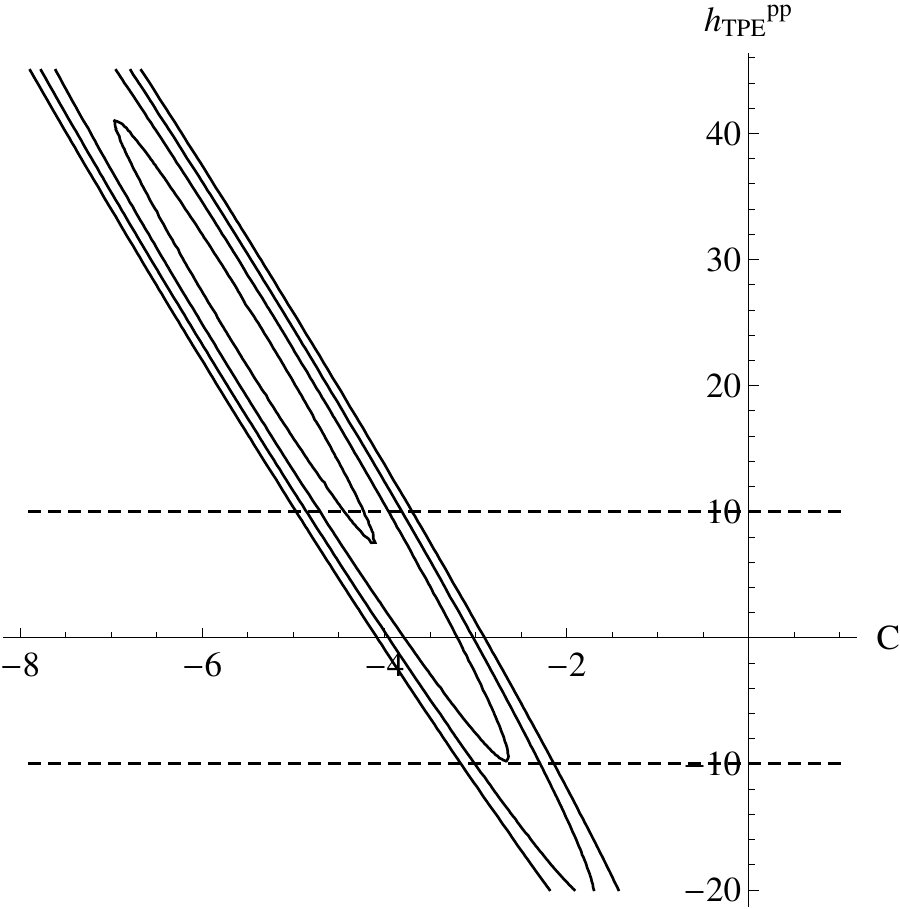}
\caption{Contours of total $\chi^2=1,2,3,4$ ($\chi^2_\mathrm{min}\simeq 0.6$ for $1$ d.o.f.)  in the $C$-$\bar h_{\mathrm{TPE}}^{pp}$ plane ($C$ in units of $10^{-6}$ and $\bar h_{\mathrm{TPE}}^{pp}$ in $10^{-6}\,\mathrm{GeV}^{-1}$). The dashed lines indicate a bound on $\bar h^{pp}_{\mathrm{TPE}}$ from dimensional analysis and the perturbativeness of the chiral expansion. }
 \label{Fithbar}
\end{figure}

We can study this in a bit more detail by choosing $h_\pi = 1\cdot 10^{-7}$ (the lattice value \cite{hpilatt}) and fitting $C$ and $\bar h^{pp}_{\mathrm{TPE}}$ to the $pp$ LAP data. In Fig.~\ref{Fithbar} we show contours of a constant total  $\chi^2=1,2,3,4$ in the $C$-$\bar h_{\mathrm{TPE}}^{pp}$ plane. The range for $C$ is similar to the range obtained in Eq.~\eqref{fitNLOplus}, and the best fit for the N${}^2$LO LEC is $\bar h^{pp}_{\mathrm{TPE}} \simeq 2.4\cdot 10^{-5}\,\mathrm{GeV}^{-1}$.

Such a value of $\bar h^{pp}_{\mathrm{TPE}}$ would be problematic. First of all, at such a large value the contributions from the NLO  and N${}^2$LO potential are of the same size. That is, the chiral expansion breaks down due to the unnaturally large size of $\bar h^{pp}_{\mathrm{TPE}}$. This can also be seen by comparing the best fit value to the dimensional estimate in Eq.~\eqref{NDA2} which predicts \mbox{$\bar h^{pp}_{\mathrm{TPE}} \simeq 1\cdot 10^{-6}\,\mathrm{GeV}^{-1}$}. Of course, such a dimensional estimate is not very precise and dimensionless factors\footnote{Note, for example, the factor $5 g_A^2/4\simeq 4$ appearing in the expression of $\bar h^{pp}_{\mathrm{TPE}}$ in Eq.~\eqref{defhbar}.} could shift the expectations. Nevertheless, it is not expected that the LECs are larger than, say, a factor $10$ than dimensional analysis suggests. The dashed horizontal lines in Fig.~\ref{Fithbar} denote this bound. If $\bar h^{pp}_{\mathrm{TPE}}$ takes on a value within the bounded region, the N${}^2$LO corrections are at most $40\%$ of the NLO contributions which is of the expected size. This would imply
 \begin{eqnarray}\label{fitN2LO}
h_\pi = 1\cdot 10^{-7}\,,\qquad
C = (-3.5\pm 1.5)\cdot 10^{-6}\,,\qquad
\bar h^{pp}_{\mathrm{TPE}} = (0\pm1)\cdot 10^{-5}\,\mathrm{GeV}^{-1}\,\,\,,
\end{eqnarray}
and there is not enough data to make a more precise statement regarding the size of $\bar h^{pp}_{\mathrm{TPE}}$. If $ h_\pi$ takes on such a small value, the contributions proportional to $h_\pi$ are about $10\%$ of the contributions proportional to $C$.

Although the above considerations where to some extend based on dimensional arguments which are unreliable, we can still draw a number of conclusions. First of all, the N${}^2$LO OPE corrections are most likely much smaller than N${}^2$LO TPE corrections as indicated by Fig.~\ref{Az_N2LO_TPE}. Secondly, unless $h_\pi$ is very small (or the N${}^2$LO LECs are very big), the N${}^2$LO TPE corrections are dominated by terms proportional to $h_\pi$ which can be seen by comparing the blue (solid) and brown (dotted) lines in the left-panel of Fig.~\ref{Az_N2LO_TPE}. This would imply that the dominant part of the N${}^2$LO potential contains no new LECs. Finally, if it turns out that $|h_\pi| \lesssim 10^{-7}$, N${}^2$LO corrections proportional to new LECs might need to be included. However, in this case the contact terms dominate the potential and the N${}^2$LO LECs need to be larger than expected by dimensional analysis in order to play a noticeable role.

Only additional data on $\slashP$ hadronic processes can tell us which of the above scenarios is realized in nature. Nevertheless, for now it seems safe\footnote{As has been said before, the $pp$ LAP is somewhat special since the LO $\slashP$ potential does not contribute. In more general processes where it does contribute, the N${}^2$LO corrections are expected to be even less important.} to neglect the N${}^2$LO corrections when describing hadronic $P$ violation. A possible exception is the correction in Eq.~\eqref{c4} which, however, brings in no new LECs. This means that the potential still depends on a total of six LECs consisting of $h_\pi$ and five contact terms. If future data cannot be described in terms of these six LECs, the remaining N${}^2$LO corrections calculated in this work might need to be included.

\section{Conclusions}\label{conclusion}
Most studies of hadronic flavor-conserving $P$ violation have used the one-meson exchange model of Ref.~\cite{Desplanques:1979hn} to describe the available data. In this so-called DDH model hadronic $P$ violation is described in terms of seven meson-nucleon coupling constants. Starting with the work of Ref.~\cite{Kaplan:1992vj}, a more systematic approach based on chiral symmetry considerations has been investigated as well. This has lead to a derivation of the $\slashP$ chiral $N\!N$ potential \cite{Zhu,KaiserPodd,Girlanda:2008ts} up to next-to-leading order. This potential consists of six LECs consisting of the pion-nucleon vertex $h_\pi$ and five $N\!N$ contact terms. The pion-nucleon vertex not only gives rise to a one-pion-exchange potential, which appears in the DDH model as well, but also to two-pion-exchange contributions with a nontrivial dependence on the exchanged momentum.

It is useful to compare calculations of hadronic $\slashP$ processes between the different frameworks. In this work we have used resonance saturation techniques~\cite{ Epelbaum:2001fm, Berengut:2013nh} to construct a dictionary between parameters appearing in the DDH and chiral potentials. The contact interactions appearing in the latter are described by the single exchange of a $\rho$- or $\omega$-meson in the DDH model. However, care must be taken for the two-pion-exchange contributions appearing in the chiral potential. We have derived explicit relations which take this subtlety into account. We have used these relations to predict the values of the contact LECs $C_i$ using model calculations of the DDH parameters. We find that the predictions of the $C_i$ are of natural size.

In addition, we have compared calculations of the proton-proton longitudinal analyzing power in both approaches \cite{Carlson:2001ma,Holsteinreview, deVries}. The relations obtained in this work can be used to translate the extracted DDH parameters into the two relevant LECs ($h_\pi$ and $C$). We have shown, see Fig.~\ref{DDH_EFT_1}, that the two approaches give consistent results. Nevertheless, we stress that the chiral approach has certain advantages over the DDH model. In particular, the former can be systematically extended to other processes (\textit{e.g.} electromagnetic processes) and higher-order corrections can be calculated.

Motivated by the possible smallness of the leading-order pion-nucleon coupling $h_\pi$, we have calculated and investigated next-to-next-to-leading-order (N${}^2$LO) corrections to the $\slashP$ $N\!N$ potential. We have applied the same power counting as used for the successful construction of the $P$-even potential \cite{Epelbaum:2004fk}. In total, the N${}^2$LO one- and two-pion-exchange-corrections depend on five new LECs, in addition to a two-pion-exchange correction proportional to $h_\pi c_4$, where $c_4\simeq 3.4$ GeV${}^{-1}$ is a strong correction to the two-pion-nucleon vertex.

Because $c_4$ is rather large, we have first investigated the impact of the $h_\pi c_4$-correction on $pp$ scattering. In Sect.~\ref{refitc4}, we have extracted new values for the LECs $h_\pi$ and $C$ taking the new correction into account. We find that the extraction is affected by approximately $20\%$ consistent with the power-counting expectations. The new values are
 \begin{eqnarray}
h_\pi &=& (0.8\pm 1.5 )\cdot 10^{-6}\,\,\,,\nonumber\\
C &=& (-5.5\pm 7)\cdot 10^{-6}\,\,\,,\nonumber
\end{eqnarray}
and fall within the error margins of the NLO extraction of Ref.~\cite{deVries}. We conclude that the power counting is working satisfactorily. The error margins are completely dominated by the lack of data. The theoretical uncertainty due to varying the cut-off parameters and missing higher-order effects is only roughly $10\%-20\%$ of the experimental one, see Sects.~\ref{refitc4} and \ref{remainder} and Ref.~\cite{deVries} for more details.

In the next step we have investigated the impact of the remaining N${}^2$LO corrections which depend on new, unknown LECs. In particular, the $pp$ longitudinal analyzing power depends on two combinations of new LECs: $\bar h^{pp}_{\mathrm{OPE}}$ (due to one-pion exchange) and $\bar h^{pp}_{\mathrm{TPE}}$ (due to two-pion exchange).
By assuming $\bar h^{pp}_{\mathrm{OPE}}$ and $\bar h^{pp}_{\mathrm{TPE}}$ to be of similar size, we show that the OPE corrections can be safely neglected since they are significantly smaller than the TPE corrections, possibly due to the momentum dependence of the former. In addition, a calculation of the $pp$ longitudinal analyzing power shows that unless $h_\pi$ is highly suppressed, the N${}^2$LO $h_\pi c_4$-correction dominates the $\bar h^{pp}_{\mathrm{TPE}}$ terms. We conclude that only if $|h_\pi| \lesssim 10^{-7}$, the $\bar h^{pp}_{\mathrm{TPE}}$ should be included in the analysis.

In summary, we have investigated the parity-violating nucleon-nucleon potential. We have derived relations between the six LECs appearing in the next-to-leading-order potential to the seven parameters appearing in the DDH one-meson exchange potential. In addition, we have calculated, for the first time, the next-to-next-to-leading-order corrections to the parity-odd potential. We conclude that, unless the leading order weak pion-nucleon vertex is highly suppressed, the next-to-next-to-leading-order corrections can be neglected. More data is needed in order to make firmer statements. If future experimental results cannot be described by the six LECs in the NLO potential, the higher-order corrections calculated here should be included in the analysis.

\subsection*{Acknowledgements}

We are grateful to Andreas Nogga and Evgeny Epelbaum for useful
discussions. This work is supported in part by the DFG and the NSFC
through funds provided to the Sino-German CRC 110 ``Symmetries and
the Emergence of Structure in QCD'' (Grant No. 11261130311).
We acknowledge the support of the European Community-Research Infrastructure 
Integrating Activity ``Study of Strongly Interacting Matter'' (acronym
HadronPhysics3, Grant Agreement n. 283286) under the Seventh Framework Programme of EU.

\appendix
\section{Resonance saturation relations}\label{AppA}
The relations between the chiral NLO LECs $h_\pi$ and $C_i$ for $S\leftrightarrow P$ transitions are given in Eqs.~\eqref{comparison} and \eqref{TPEshift}. Here we give similar relations for the $P\leftrightarrow D$ transitions. Since both the LO chiral potential and the DDH potential contain the OPE contribution in Eq.~\eqref{onepion}, OPE terms are not included in the relations below. If a regulator function is applied to the OPE potential in the DDH framework, the relations below should be modified accordingly.

In contrast to the DDH potential in Eq.~\eqref{DDHpot}, the NLO TPE potential in Eq.~\eqref{NLOTPE} does not contribute to all $P\leftrightarrow D$ transitions. The non-vanishing transitions are the ${}^3P_2\leftrightarrow {}^1D_2$ transition for $m_t = \pm 1$  ($m_t$ is the third component of the total isospin of the interacting nucleon pair) while for $m_t=0$ the chiral potential does not contribute. This gives the following relation
\begin{equation}
 \frac{m_t \,g_A^3h_{\pi}}{\sqrt{2}F_{\pi}(4\pi F_{\pi})^2m_{\pi}^2}\frac{s^3}{\Lambda_S^3}\sim \frac{-3}{m_N}\left[\frac{g_{\omega}\chi_S}{m_{\omega}^4}(h_{\omega}^0+ m_t h_{\omega}^1)c_{\omega}^{\prime}+\frac{g_{\rho}\chi_V}{m_{\rho}^4}\left(h_{\rho}^0+ m_t h_{\rho}^1-(2-3 m_t^2)\frac{h_{\rho}^2}{\sqrt{6}}\right)c_{\rho}^{\prime}\right]\,,
\end{equation}
where $c_{\rho,\omega}^{\prime}$ is defined in Eq.~\eqref{cprime}.
Both potentials contribute to ${}^3P_1\leftrightarrow {}^3D_1$ and  ${}^3P_2\leftrightarrow {}^3D_2$ transitions
\begin{equation}
\frac{g_Ah_{\pi}}{4\sqrt{2}F_{\pi}(4\pi F_{\pi})^2m_{\pi}^2}\frac{s\left[\Lambda_S^2-4\mpi^2 + g_A^2(8\mpi^2-5\Lambda_S^2)\right]}{\Lambda_S^3}
\sim\frac{-3}{m_N}\left[\frac{g_{\rho}}{m_{\rho}^4}(h_{\rho}^1+h_{\rho}^{1\prime})c_{\rho}^{\prime}-\frac{g_{\omega}}{m_{\omega}^4}h_{\omega}^1c_{\omega}^{\prime}\right]\,.
\end{equation}
The ${}^1P_1\leftrightarrow {}^3D_1$ transition does not get a TPE contribution such that
\begin{equation}
0\sim \frac{g_{\omega}}{m_{\omega}^4}h_{\omega}^0 c_{\omega}^{\prime}(2+\chi_S)-\frac{3g_{\rho}}{m_{\rho}^4}h_{\rho}^0 c_{\rho}^{\prime}(2+\chi_V).
\end{equation}

It is clear that the $P \leftrightarrow D$ transitions are much more constrained in the NLO chiral potential where they only depend on a single LEC, compared to the DDH model in which the transitions depend on five different combinations of  parameters.

\end{document}